\documentclass{ws-procs9x6}
\begin{document}

\title{Fermionisation of a Two-Dimensional\\
Free Massless Complex Scalar Field}

\author{Laure GOUBA$^{1,2}$, Gabriel Y.H. AVOSSEVOU$^{1,2}$,\\
Jan GOVAERTS$^3$ and M. Norbert HOUNKONNOU$^{1,2}$}

\address{$^1$International Chair in Mathematical Physics 
and Applications (ICMPA)\\
072 B.P. 50, Cotonou, Republic of Benin}

\address{$^2$Unit\'e de Recherche en Physique Th\'eorique (URPT)\\
Institut de Math\'ematiques et de Sciences Physiques (IMSP)\\
B.P. 2628 Porto-Novo 01, Republic of Benin\\
E-mail: laure\_gouba@cipma.net, avossevou@yahoo.fr,
norbert\_hounkonnou@cipma.net}

\address{$^3$Institute of Nuclear Physics,
Catholic University of Louvain\\
2, Chemin du Cyclotron, B-1348 Louvain-la-Neuve, Belgium\\
E-mail: govaerts@fynu.ucl.ac.be}


\maketitle

\abstracts{The fermionisation of a two-dimensional free
massless complex scalar field is given through its
derivative field which is a conformal field.}

\section{Introduction}

Recently,\cite{ppz} it has been shown that the massless 
Schwinger model, namely two-dimensional massless
spinor electrodynamics, may be solved exactly 
through the bosonisation of the massless fermion
without having recourse to any gauge fixing procedure.\cite{ja}
With this background in mind, and the possible resolution of 
massless scalar electrodynamics in two-dimensional Minkowski 
spacetime compactified into the cylinder geometry for the space coordinate, 
in this communication we discuss the fermionisation of
a massless complex scalar field. As is the case for the Schwinger
model, one may expect that this could lead to nonlinear
realisations of gauge symmetries for fermionic fields.

Fermionisation is the process of writing
bosonic variables in terms of fermionic ones,
with the aim here to solve exactly the $2$-dimensional scalar
electrodynamics model. The model we are interested in is the
free massless complex scalar field defined as
\begin{equation}\label{1}
\phi = \frac{1}{\sqrt 2}\left( \varphi + i\chi\right),
\end{equation}
where $\varphi$ and $\chi$ are, respectively, the real
and the imaginary parts of $\phi$.

\section{Fermionization of a Real Scalar Field}

Let us consider the free massless real scalar field $\varphi(t,x)$
over spacetime, with a dynamics governed by the local spacetime action
\begin{equation}
S[\varphi] =\int d^2x^\mu {\mathcal L}
(\varphi,\partial_\mu\varphi)\ ,
\end{equation}
and the Lagrangian density
\begin{equation}
 {\mathcal L}
=\frac{1}{2}(\partial_\mu\varphi)^2\ .
\end{equation}
As usual, the spacetime coordinate indices take the values
$\mu=(0,1)$, while the Minkowski spacetime metric signature is
${\rm diag}\,\eta_{\mu\nu}=(+-)$. We also assume a system of units
such that $c=1=\hbar$.
In their manifestly Lorentz covariant form,
the Euler--Lagrange equations read
\begin{equation}
\partial_\mu\frac{\partial{\mathcal L}}{\partial
(\partial_\mu\varphi)}-
\frac{\partial{\mathcal L}}{\partial\varphi}=0\ ,
\end{equation}
reducing, in the present case, to the massless Klein--Gordon equation,
\begin{equation}
(\partial_0^2 - \partial_1^2)\varphi(t,x) = 0\ .
\end{equation}
Given periodic boundary conditions associated to the circle geometry
of the space coordinate $x$ with period $L$, through direct discrete
Fourier analysis the general solution is readily established as follows,
\begin{eqnarray}\nonumber
\varphi(t,x) &=& \frac{1}{\sqrt{4\pi}}\left\{ q_0 +
\frac{4\pi}{L}\alpha_0t + i\sum_{n\ge 1}
\left( \frac{1}{n}\alpha_ne^{-\frac{2i\pi}{L}n(t+x)}
-\frac{1}{n}\alpha^{\dagger}_ne^{\frac{2i\pi}{L}n(t+x)}\right)
\right.\\
&&+ \left. i\sum_{n \ge 1}\left(
\frac{1}{n}\bar{\alpha}_n e^{-\frac{2i\pi}{L}n(t-x)}
- \frac{1}{n}\bar{\alpha}_n^{\dagger}e^{\frac{2i\pi}{L}n(t-x)}
\right)\right\}\ .
\end{eqnarray}
By definition, the momentum conjugate to the field
$\varphi(t,x)$ at each point $x$ in space is
\begin{equation}
\pi_{\varphi} = 
\frac{\partial{\mathcal L}}{\partial (\partial_0\varphi)}=
\partial_0\varphi\ ,
\end{equation}
leading, in the present case, to the mode expansion,
\begin{eqnarray}\nonumber
\pi_{\varphi}(t,x) &=& \frac{1}{\sqrt{4\pi}}\left\{ \frac{4\pi}{L}
\alpha_0 + \frac{2\pi}{L}\sum_{n\ge 1}\left(\alpha_n
e^{-\frac{2i\pi}{L}n(t+x)} + \alpha_n^\dagger
e^{\frac{2i\pi}{L}n(t+x)}\right)\right.\\
&&+\left. \frac{2\pi}{L}\sum_{n\ge 1}\left( \bar{\alpha}_n
e^{-\frac{2i\pi}{L}n(t-x)} + \bar{\alpha}^\dagger_n
e^{\frac{2i\pi}{L}n(t-x)}\right)\right\}\ ,
\end{eqnarray}
such that the corresponding  nonvanishing Poisson brackets read
\begin{eqnarray}
\left\{ q_0,\alpha_0\right\}= 1\ \ \ ,\ \ \
\left\{\alpha_n,\alpha^\dagger_m\right\} = -in\delta_{n,m}
= \left\{\bar{\alpha}_n,\bar{\alpha}^\dagger_m\right\}\ ,
\end{eqnarray}
\begin{eqnarray}
\left\{ \varphi(t,x), \pi(t,y)\right\}
= \frac{1}{L}\sum_{n= -\infty}^{+\infty}
e^{-\frac{2i\pi}{L}n(x-y)}
= \sum_{n= -\infty}^{+\infty}\delta(x-y+nL)\ .
\end{eqnarray}

Next, the quantum analogue of the classical field, expressed in the
Schr\"odinger picture at $t=0$, writes
\begin{eqnarray}\nonumber
\varphi(x) &=& \frac{1}{\sqrt{4\pi}}\left[ q_0 
+ \frac{4\pi}{L}\alpha_0\frac{x-x}{2}
+ i\sum_{n\ge 1}\frac{1}{n}\left(\alpha_n 
e^{-\frac{2i\pi}{L} n x}-
\alpha_n^\dagger e^{\frac{2i\pi}{L}n x}\right)\right.
\\ &&+\left.
i\sum_{n\ge 1}\frac{1}{n}\left(
\bar{\alpha}_n e^{\frac{2i\pi}{L}n x} -
\bar{\alpha}_n^\dagger
e^{-\frac{2i\pi}{L}n x}\right)\right]\ ,
\end{eqnarray}
with the corresponding nonvanishing commutators,
\begin{equation}
\left[q_0,\alpha_0\right] = i\ \ ,\ \ 
\left[ \alpha_n,\alpha^\dagger_m\right]=n\delta_{n,m}\ \ ,\ \ 
\left[ \bar{\alpha}_n, \bar{\alpha}_m^\dagger\right]=n\delta_{n,m}\ .
\end{equation}

Furthermore, the chiral decomposition of 
$\varphi(x)=\varphi_+(x)+\varphi_-(x)$ is given by
\begin{eqnarray}\nonumber
\varphi_{\pm }(x) &=& \frac{1}{\sqrt{4\pi}}\left\{ q_{\pm ,0} +
\frac{2\pi}{L}\alpha_{\pm ,0}
(\pm x)\right.\\
&&+\left. i\sum_{n\ge 1}\left( \frac{1}{n}
\alpha_{\pm ,n}
e^{-\frac{2i\pi}{L}n(\pm x)} - 
\frac{1}{n}\alpha^\dagger_{\pm ,n}
e^{\frac{2i\pi}{L}n( \pm x)}\right) \right\}\ ,
\end{eqnarray}
with the identifications,
 \begin{equation}
q_{\pm,0} =\frac{1}{2}q_0\ \ ,\ \ 
\alpha_{\pm,0}=\alpha_0\ \ ,\ \ 
\alpha_{+,n}=\alpha_n\ \ ,\ \ 
\alpha_{-,n}=\bar{\alpha}_n\ ,
\end{equation}
and the nonvanishing commutators:
\begin{equation}
\left[\alpha_{+,n},\alpha^\dagger_{+,m}\right] 
= n\delta_{n,m}= \left[\alpha_{-,n},\alpha^\dagger_{-,m}\right]\ .
\end{equation}

In the sequel, we introduce the change of variable to the complex plane
\begin{equation}
z=e^{\frac{2i\pi}{L}(\pm x)}\ \ \ ,\ \ \ 
\pm x =-i\frac{L}{2\pi}\ln(z)\ ,
\end{equation}
so that
\begin{equation}
\varphi_{\pm}(z) = \frac{1}{\sqrt{4\pi}}\left[q_{\pm,0}-
i\alpha_{\pm,0} \ln(z)+i \sum_{n\ge 1}\left(\frac{1}{n}\alpha_{\pm,n}z^{-n} 
- \frac{1}{n}\alpha_{\pm,n}^\dagger z^n\right)\right]\ .
\end{equation}
It is worth noticing that the field $\varphi_{\pm}(z)$ is not 
a conformal field. Rather, its  derivative is such a conformal field
of weight unity,
\begin{equation}
\partial_z\varphi_{\pm}(z) = \frac{1}{\sqrt{4\pi}}\left[
-i\alpha_{\pm,0} \frac{1}{z} - i \sum_{n\ge 1}
\left(\alpha_{\pm,n}z^{-n-1} + \alpha_{\pm,n}^\dagger z^{n-1}
\right)\right]\ .
\label{2}
\end{equation}
Then from (\ref{2}), we get
\begin{equation}
i\sqrt{4\pi}\partial_z\varphi_{\pm}(z)= \alpha_{\pm,0}
\frac{1}{z} + \sum_{n\ge 1}\left(\alpha_{\pm,n}
z^{-n-1}+\alpha_{\pm,n}^\dagger z^{n-1}\right)\ .
\end{equation}
Consequently, the modes may be given the following
contour representations in the complex plane,
\begin{eqnarray}
\alpha_{\pm,n} &=& \oint_0\frac{dz}{2i\pi}\left(
i\sqrt{4\pi}\partial_z\varphi_{\pm}(z)  \right)z^n\ ,\\
\alpha_{\pm,n}^\dagger &=& \oint_0\frac{dz}{2i\pi}\left(
i\sqrt{4\pi}\partial_z\varphi_{\pm}(z)  \right)z^{-n}\ ,\\
\alpha_{\pm,0} &=& \oint_0\frac{dz}{2i\pi}\left(i\sqrt{4\pi}
\partial_z\varphi_{\pm}(z)\right)\ ,
\end{eqnarray}
where the contour $\oint_0$ is taken around the origin $z=0$.

\vspace{10pt}

Next, in order to introduce the fermionic degrees
of freedom towards the fermionisation of the bosonic field, 
let us consider the $2$-dimensional free massless fermionic 
Dirac field $\psi(t,x)$ with the corresponding Lagrangian density
\begin{equation}
{\mathcal{L}} =i\,\bar\psi\gamma^\mu\partial_\mu\psi\ ,
\end{equation}
with $\bar\psi = \psi^\dagger\gamma^0$, $\psi^\dagger$ being
the Hermitian conjugate of the bi-spinor $\psi$. The matrices $\gamma^\mu$
define the associated Clifford--Dirac algebra in two-dimensional Minkowski
spacetime, a representation of which is provided by the Pauli matrices,
\begin{eqnarray}
\gamma^0 = \left(
\begin{array}{cc}
0 & 1\\
1 & 0
\end{array}\right)\ \ \ ,\ \ \ 
\gamma^1 = \left(
\begin{array}{cc}
0 & 1\\
-1 & 0
\end{array}\right)\ \ \ ,\ \ \ 
\gamma_5=\gamma^0\gamma^1= \left(
\begin{array}{cc}
-1 & 0\\
0 & 1
\end{array}\right)\ ,
\end{eqnarray}
while the Dirac spinor in that representation decomposes according to,
\begin{eqnarray}\label{6}
\psi =\left(
\begin{array}{cc}
\psi_+\\
\psi_-
\end{array}\right)\ \ \ ,\ \ \ 
\psi^\dagger =\left( \psi^\dagger_+\ \ \psi^\dagger_-\right)\ ,
\end{eqnarray}
where $\psi_{\pm}$ are the chiral components of the field $\psi$.

The equations of motion, in this case, read
\begin{equation}
\partial_0\psi_{\pm} = \pm\partial_1\psi_{\pm}.
\end{equation}
Assuming again periodic boundary conditions in $x$ for the Dirac field
(more general choices are also possible),
the solutions to these equations may be expressed through the discrete
Fourier mode decomposition,
\begin{equation}\label{3}
\psi_{\pm}(t,x) = \frac{1}{\sqrt{L}}\sum_{n=-\infty}^{+\infty}\psi_{\pm,n}
e^{-\frac{2i\pi}{L}n(t\pm x)}\ .
\end{equation}

Once again at the quantum level and in the Sch\"odinger picture at $t=0$, 
substituting for the change of variable
\begin{equation}\label{5}
z=e^{\frac{2i\pi}{L}(\pm x)}
\end{equation}
in (\ref{3}), one obtains,
\begin{equation}
\sqrt{L}\ \psi_{\pm}(z) =
\sum_{n=-\infty}^{+\infty}\psi_{\pm,n}z^{-n}\ \ \ ,\ \ \ 
\sqrt{L}\ \psi_{\pm}^{\dagger}(z) =
\sum_{n=-\infty}^{+\infty}\psi_{\pm,n}^{\dagger}z^n\ .
\end{equation}
The fermionic modes $\psi_{\pm,n}$ obey the following anticommutation
relations,
\begin{equation}
\left\{\psi_{+,n},\psi^\dagger_{+,m}\right\}=\delta_{n,m}=
\left\{\psi_{-,n},\psi^\dagger_{-,m}\right\}\ ,
\end{equation}
so that for the fields themselves,
\begin{equation}
\left\{\psi(x),\psi^\dagger(y)\right\}=
\sum_{n=-\infty}^{+\infty}\delta(x-y+nL)\ .
\end{equation}
The set of independent bilinear fermionic currents are in fact, in the present
situation,
%\begin{equation}
$\overline{\psi}\gamma^\mu\psi$,
%\end{equation}
namely,
\begin{equation}
\overline{\psi}\gamma^0\psi=\psi^\dagger_+\psi_++\psi^\dagger_-\psi_-\ \ ,\ \ 
\overline{\psi}\gamma^1\psi=-\psi^\dagger_+\psi_++\psi^\dagger_-\psi_-\ .
\end{equation}

Let us now introduce the following association of fields, where in the r.h.s.
the double-dot notation stands for usual normal-ordering with respect
to the fermionic modes $\psi_{\pm,n}$,
\begin{equation}
\left(i\sqrt{4\pi}\partial_z\varphi_{\pm}(z)\right)
=L : \psi_{\pm}(z)\psi^{\dagger}_{\pm}(z):\ .
\label{eq:fermion}
\end{equation}
In particular, the normal ordered product for the fermionic zero modes
is taken to be,
\begin{equation}
:\psi_{\pm,0}\psi^\dagger_{\pm,0}:\ =\ \frac{1}{2}
\left[\psi_{\pm,0}\psi^\dagger_{\pm,0}-\psi^\dagger_{\pm,0}\psi_{\pm,0}\right]
\ .
\end{equation}

That the association (\ref{eq:fermion})
is meaningful follows from the algebra realised
for the modes of the field combinations on the r.h.s., which indeed is 
that of the bosonic modes associated to the field on the l.h.s. of this
correspondence as shall now be established. Consider the modes
\vspace{5pt}
\begin{equation}
\alpha_{\pm,n}=\oint_0\frac{dz}{2i\pi}(i\sqrt{4\pi}\partial_z
\varphi_{\pm}(z))z^n =L \oint_0\frac{dz}{2i\pi}:\psi_{\pm}(z)
\psi^\dagger_{\pm}(z):z^n\ ,
\end{equation}
\begin{equation}
\alpha_{\pm,m}^\dagger=\oint_0\frac{dz}{2i\pi}(i\sqrt{4\pi}\partial_z
\varphi_{\pm}(z))z^{-m}=L \oint_0\frac{dz}{2i\pi}:
\psi_{\pm}(z)\psi_{\pm}^\dagger(z):z^{-m}\ .
\end{equation}
\begin{equation}
\alpha_{\pm,0}=\oint_0\frac{dz}{2i\pi}(i\sqrt{4\pi}\partial_z
\varphi_{\pm}(z)) =L \oint_0\frac{dz}{2i\pi}:\psi_{\pm}(z)
\psi^\dagger_{\pm}(z):\ ,
\end{equation}
\vspace{5pt}
of which the commutators are given by,
\begin{equation}
\left[\alpha_{\pm,n},\alpha_{\pm,m}^\dagger\right] =
L^2\oint_0 \frac{dx}{2i\pi}x^n\oint_0 \frac{dy}{2i\pi}y^{-m}
\left[:\psi_{\pm}(x)\psi^{\dagger}_{\pm}(x):\,,
:\psi_{\pm}(y)\psi_{\pm}^\dagger(y):\right]\ .
\label{eq:commutator}
\end{equation}

Using well established techniques from conformal field theory,\cite{bu}
based on Wick's theorem relating (anti)commutators of normal-ordered
products to radial-ordered (R-ordered) products of conformal fields in
the complex plane, the explicit evaluation of the above commutator and 
contour integrations is standard, starting with the R-ordered product of 
the fermionic fields,
\begin{eqnarray}\nonumber
R\left(\psi_\pm(x)\psi^\dagger_\pm(y)\right)&=&
\frac{1}{2L}\,\frac{1}{x-y}
\left[\,\sqrt{\frac{x}{y}}+\sqrt{\frac{y}{x}}\,\right]\ +\
:\psi_\pm(x)\psi^\dagger_\pm(y):\\
&=& \frac{1}{L} \frac{1}{x-y}\ +\ \cdots\ ,
\end{eqnarray}
where the unspecified terms are regular as $x$ and $y$ approach one another
in the complex plane. Applying this result to the above relevant products, 
one then obtains,
\begin{equation}
R\left(:\psi_{\pm}(x)\psi^{\dagger}_{\pm}(x):
:\psi_{\pm}(y)\psi^\dagger_{\pm}(y):\right)=
\frac{1}{L^2}\frac{1}{(x-y)^2}\ +\ \cdots\ .
\end{equation} 

By appropriately choosing the integration contours in (\ref{eq:commutator})
in order to replace the commutator by the R-ordered product of the
two normal-ordered fermion bilinears $:\psi_\pm(x)\psi^\dagger_\pm(x):$
and $:\psi_\pm(y)\psi^\dagger_\pm(y):$, namely,
\begin{eqnarray}
&&\left[\alpha_{\pm,n},\,\alpha^\dagger_{\pm,m}\right]= \\
\nonumber
&=&L^2\oint_{|x|>|y|-|y|>|x|}\frac{dx}{2i\pi}\frac{dy}{2i\pi}x^ny^{-m}\,
R\left(:\psi_\pm(x)\psi^\dagger_\pm(x)::\psi_\pm(y)\psi^\dagger_\pm(y):\right)
\ ,
\end{eqnarray}
where $\oint_{|x|>|y|-|y|>|x|}$ stands for a contour integration combining
the difference of two double contours around the origin in both $x$ and $y$
such that $|x|>|y|$ for the first contribution and $|y|>|x|$ for
the second, and then deforming the combined contour for the variable $x$, 
given a fixed value for $y$, one has
\begin{equation}
\left[\alpha_{\pm,n}\,,\,\alpha^\dagger_{\pm,m}\right]=
\oint_0\frac{dy}{2i\pi}y^{-m}\oint_y\frac{dx}{2i\pi}
\frac{x^n}{(x-y)^2}\ ,
\end{equation}
where the second contour $\oint_y$ in $x$ is now taken around $y$.
Completing the last two contour integrations, one then readily finds,
\begin{equation}
\left[\alpha_{\pm,n}\,,\,\alpha^\dagger_{\pm,m}\right]=n\delta_{n,m}\ ,
\end{equation}
reproducing indeed the bosonic mode algebra for the chiral fields
$\partial_z\varphi_\pm(z)$ as announced.

In a likewise manner, it may be shown that all the other relevant
bosonic mode commutation relations, inclusive of those for the
zero modes $\alpha_{\pm,0}$ are also reproduced. In particular,
because of the masslessness of the fermion leading to
chirality conservation laws, all the modes of opposite chiralities
have identically vanishing commutators.
 
Note~that~the~fermion/boson~correspondence~$\left(
i\sqrt{4\pi}\partial_z\varphi_\pm(z)\right)=
\left(L:\psi_\pm(x)\psi^\dagger_\pm(z):\right)$ is in fact encapsulated
in the short distance operator product expansion (OPE)
for the following R-ordered products,
\begin{equation}
R\left(\left(L:\psi_\pm(x)\psi^\dagger_\pm(x):\right)\,
\left(L:\psi_\pm(y)\psi^\dagger_\pm(y):\right)\right)=
\frac{1}{(x-y)^2}\ +\ \cdots\ ,
\end{equation}
\begin{equation}
R\left(\left(i\sqrt{4\pi}\partial_x\varphi_\pm(x)\right)\,
\left(i\sqrt{4\pi}\partial_y\varphi_\pm(y)\right)\right)=
\frac{1}{(x-y)^2}\ +\ \cdots\ .
\end{equation}

That the above modes of the chiral conformal fields 
$L:\psi_\pm(z)\psi^\dagger_\pm(z):$ are indeed bosonic modes in $1+1$
dimension thus follows from their commutation relations obtained from
the fundamental fermionic algebra or from the associated OPE's. 
In other words, the representation
space of quantum states for the fermionic fields
$\psi_{\pm}(x)$ also provides a representation space for the bosonic algebra
associated to the conformal field $i\sqrt{4\pi}\partial_z\varphi_\pm(z)$. 
Hence, the derivative of the real scalar field may be fermionised
in this manner in terms of the above massless fermionic field
degrees of freedom in two-dimensional Minkowski spacetime. Note that
the only mode of the real scalar field $\varphi(t,x)$ itself which is not 
yet accounted for is the constant zero-mode $q_0=2q_{\pm,0}$, while the
restriction $\alpha_{+,0}=\alpha_{-,0}$ still needs to be enforced as well.

\section{Fermionization of the Imaginary Part}

The analysis for the imaginary part $\chi(t,x)$ of the complex
scalar field proceeds obviously along the same lines as those
detailed in the previous section. For the mode expansion, one has,
\begin{eqnarray}\nonumber
\chi(t,x) &=& \frac{1}{\sqrt{4\pi}}\left\{ p_0
 + \frac{4\pi}{L}\beta_0t + i\sum_{n\ge 1}\left( \frac{1}{n}\beta_n
e^{-\frac{2i\pi}{L}n(t+x)}-\frac{1}{n} 
\beta^{\dagger}_n e^{\frac{2i\pi}{L}n(t+x)}\right)\right.\\
&&+\left. i\sum_{n \ge 1}\left(
\frac{1}{n}\bar{\beta}_n e^{-\frac{2i\pi}{L}n(t-x)}-
\frac{1}{n}\bar{\beta}_n^{\dagger} e^{\frac{2i\pi}{L}n(t-x)}\right)\right\}\ ,
\end{eqnarray}
with the commutation relations,
\begin{equation}
\left[p_0,\beta_0\right]=i\ \ ,\ \ 
\left[ \beta_n,\beta^\dagger_m\right]=n\delta_{n,m}\ \ ,\ \ 
\left[ \bar{\beta}_n,\bar{\beta}_m^\dagger\right]=n\delta_{n,m}\ .
\end{equation}
The chiral components read
\begin{eqnarray}\nonumber
\chi_{\pm }(t,x) &=& \frac{1}{\sqrt{4\pi}}\left\{ p_{\pm ,0} +
\frac{2\pi}{L}\beta_{\pm ,0}
(t\pm x) \right.\\
&&+ \left.i\sum_{n\ge 1}\left( \frac{1}{n}
\beta_{\pm ,n} e^{-\frac{2i\pi}{L}n(t\pm x)}-
\frac{1}{n}\beta^{\dagger}_{\pm ,n} e^{\frac{2i\pi}{L}n(t \pm x)}
\right) \right\}\ ,
\end{eqnarray}
with the identifications
\begin{equation}
p_{\pm,0}=\frac{1}{2}p_0\ \ ,\ \ 
\beta_{\pm,0}=\beta_0\ \ ,\ \ 
\beta_{+,n}=\beta_n\ \ ,\ \ 
\beta_{-,n}=\bar{\beta}_n\ ,
\end{equation}
and the nonvanishing commutators,
\begin{equation}
\left[\beta_{+,n},\beta^\dagger_{+,m}\right]=n\delta_{n,m}=
\left[\beta_{-,n},\beta^\dagger_{-,m}\right]\ .
\end{equation}
In terms of a massless Dirac fermion field $\psi(t,x)$ considered
as in the previous section, the fermionisation of $\chi(t,x)$ is
defined by the identifications,
\begin{equation}
\beta_{\pm,n}=L\oint_0 \frac{dx}{2i\pi}
:\psi_{\pm}(x)\psi^{\dagger}_{\pm}(x): x^n\ ,
\end{equation}
\begin{equation}
\beta_{\pm,n}^\dagger=L\oint_0 \frac{dx}{2i\pi}
:\psi_{\pm}(x)\psi^\dagger_{\pm}(x): x^{-n}\ ,
\end{equation}
\begin{equation}
\beta_{\pm,0}=L\oint_0 \frac{dx}{2i\pi}
:\psi_{\pm}(x)\psi^{\dagger}_{\pm}(x):\ .
\end{equation}
The other results that have been stated for $\varphi(t,x)$
of course apply likewise for the field $\chi(t,x)$.

\section{Fermionization of the Complex Scalar Field}

It now suffices to combine the results established in the previous
two sections into the definition of the complex scalar field from
its real and imaginary parts, as given in (\ref{1}).
The mode expansion of the field $\phi(t,x)$ thus writes as follows,
\begin{eqnarray}\nonumber
\phi(t,x) &=& \frac{1}{\sqrt{4\pi}}\left\{ c_0 +
\frac{4\pi}{L}a_0t + i \sum_{n\ge 1}\left( \frac{1}{n}a_n
e^{-\frac{2i\pi}{L}n(t+x)} -
\frac{1}{n}b^{\dagger}_n e^{\frac{2i\pi}{L}n(t+x)}\right) \right.\\
&&+ \left. i\sum_{n \ge 1}\left( \frac{1}{n}\bar{a}_n
e^{-\frac{2i\pi}{L}n(t-x)} - \frac{1}{n}\bar{b}_n^{\dagger}
e^{\frac{2i\pi}{L}n(t-x)}\right)\right\}\ .
\end{eqnarray}
The modes of $\phi$ are related to those of its real and imaginary 
parts as follows,
\begin{eqnarray}
c_0 &=& \frac{1}{\sqrt{2}}(q_0 + ip_0)\ \ \ ,\ \ \ 
a_0 = \frac{1}{\sqrt{2}}(\alpha_0 + i\beta_0)\ ,\\
a_n &=& \frac{1}{\sqrt{2}}(\alpha_n+ i\beta_n)\ \ \ ,\ \ \ 
\bar{a}_n = \frac{1}{\sqrt{2}}(\bar{\alpha}_n + i\bar{\beta}_n)\ ,\\
b_n &=& \frac{1}{\sqrt{2}}(\alpha_n - i\beta_n)\ \ \ ,\ \ \ 
\bar{b}_n = \frac{1}{\sqrt{2}}(\bar{\alpha}_n - i\bar{\beta}_n)\ ,\\
a_n^\dagger &=& \frac{1}{\sqrt{2}}(\alpha_n^\dagger - i\beta_n^\dagger)
\ \ \ ,\ \ \ 
\bar{a}_n^\dagger = \frac{1}{\sqrt{2}}(\bar{\alpha}_n^\dagger - 
i\bar{\beta}_n^\dagger)\ ,\\
b_n^\dagger &=& \frac{1}{\sqrt{2}}(\alpha_n^\dagger + i\beta_n^\dagger)
\ \ \ ,\ \ \ 
\bar{b}_n^\dagger = \frac{1}{\sqrt{2}}(\bar{\alpha}_n^\dagger + 
i\bar{\beta}_n^\dagger)\ ,
\end{eqnarray}
with the nonvanishing commutation relations,
\begin{eqnarray}
\left[  a_n, a_m^\dagger\right] = n\delta_{n,m}\ \ \ &,&\ \ \ 
\left[  \bar{a}_n, \bar{a}_n^\dagger\right] = n\delta_{n,m}\ ,\\
\left[  b_n, b_m^\dagger\right] = n\delta_{n,m}\ \ \ &,&\ \ \ 
\left[  \bar{b}_n, \bar{b}_n^\dagger\right] = n\delta_{n,m}\ .
\end{eqnarray}
Collecting the chiral components of the complex field $\phi$, we get
\begin{eqnarray}\nonumber
\phi_{\pm }(t,x) &=& \frac{1}{\sqrt{4\pi}}\left\{ c_{\pm ,0} +
\frac{2\pi}{L}a_{\pm ,0}
(t\pm x)+ \right.\\
&&+ \left.i\sum_{n\ge 1}\left( \frac{1}{n} 
a_{\pm ,n} e^{-\frac{2i\pi}{L}n(t\pm x)} -
\frac{1}{n}b^{\dagger}_{\pm ,n} e^{\frac{2i\pi}{L}n(t \pm x)}
\right) \right\}\ ,
\end{eqnarray}
in a notation which should by now be obvious.

In the Schr\"odinger picture at $t=0$, and using the change of
variable (\ref{5}), the field may be expressed as,
\begin{equation}
\phi_{\pm}(z) = \frac{1}{\sqrt{4\pi}}\left[ c_{\pm,0}
-ia_{\pm,0}\ln(z) + i \sum_{n\ge 1}\left(\frac{1}{n}
a_{\pm,n} z^{-n} - \frac{1}{n}b_{\pm,n}^\dagger z^{n}\right)\right]\ .
\end{equation}

In order to fermionise the complex scalar field $\phi(t,x)$, one needs
to introduce two massless Dirac spinors $\psi_1(t,x)$ and $\psi_2(t,x)$,
each with the mode decompositions discussed previously.
The fermionisation rule for the complex scalar field is then given by
the definitions,
\begin{eqnarray}
a_{\pm,n}= L \oint_0 \frac{dx}{2i\pi}\frac{1}{\sqrt{2}}\left[
:\psi_{1,\pm}(x)\psi^{\dagger}_{1,\pm}(x): + i
:\psi_{2,\pm}(x)\psi^{\dagger}_{2,\pm}(x):\right] x^n \ ,
\end{eqnarray}
\begin{eqnarray}
b_{\pm,n}= L \oint_0 \frac{dx}{2i\pi}\frac{1}{\sqrt{2}}\left[
:\psi_{1,\pm}(x)\psi^{\dagger}_{1,\pm}(x): - i
:\psi_{2,\pm}(x)\psi^{\dagger}_{2,\pm}(x):\right] x^n \ ,
\end{eqnarray}
\begin{eqnarray}
a^\dagger_{\pm,n}= L \oint_0 \frac{dx}{2i\pi}\frac{1}{\sqrt{2}}\left[
:\psi_{1,\pm}(x)\psi^{\dagger}_{1,\pm}(x): - i
:\psi_{2,\pm}(x)\psi^{\dagger}_{2,\pm}(x):\right] x^{-n} \ ,
\end{eqnarray}
\begin{eqnarray}
b^\dagger_{\pm,n}= L \oint_0 \frac{dx}{2i\pi}\frac{1}{\sqrt{2}}\left[
:\psi_{1,\pm}(x)\psi^{\dagger}_{1,\pm}(x): + i
:\psi_{2,\pm}(x)\psi^{\dagger}_{2,\pm}(x):\right] x^{-n} \ ,
\end{eqnarray}
\begin{eqnarray}
a_{\pm,0}= L \oint_0 \frac{dx}{2i\pi}\frac{1}{\sqrt{2}}\left[
:\psi_{1,\pm}(x)\psi^{\dagger}_{1,\pm}(x): + i
:\psi_{2,\pm}(x)\psi^{\dagger}_{2,\pm}(x):\right]\ .
\end{eqnarray}

Clearly from our previous discussion, the commutator algebra of these
different modes is indeed that of the associated complex scalar field
as described above.

\section{Final Comments}

A massless complex scalar field not being conformal, does not lend itself
to the fermionisation procedure. However, the derivative field is
a conformal field of weight unity. This latter field may easily be
fermionised in terms of two massless Dirac spinors, thereby reproducing
the non-zero mode algebra of the bosonic degrees of freedom in terms
of the fermionic algebra of the Dirac fermions. However,
fermionisation of the zero modes $c_{\pm,0}$ and $a_{\pm,0}$ with
the restriction $a_{+,0}=a_{-,0}$ remains an open question.

\section*{Acknowledgements}

The authors acknowledge the Belgian Cooperation CUD-CIUF/UAC
for financial support. LG is presently supported through a Ph.D. Fellowship
of the Third World Organisation for Women in Science (TWOWS, Third World
Academy of Science). The work of JG is partially supported by the 
Federal Office for Scientific, Technical and Cultural Affairs (Belgium) 
through the Interuniversity Attraction Pole P5/27.


\begin{thebibliography}{0}

\bibitem{ppz} 
G.Y.H. Avossevou and J. Govaerts, {\sl The Schwinger model and the physical
projector: a nonperturbative quantization without gauge fixing\/},
in the Proceedings of the Second International Workshop on 
Contemporary Problems in Mathematical Physics, J.~Govaerts, 
M.N.~Hounkonnou and A.Z.~Msezane, eds. 
(World Scientific, Singapore, 2002), pp. 374--394.  

\bibitem{ja} 
G.Y.H. Avossevou, Ph.D. Thesis (Institut de Math\'ematiques 
et de Sciences Physiques, University of Abomey-Calavi,
Republic of Benin, April 2002), unpublished.

\bibitem{bu}
Paul Ginsparg, {\sl Applied conformal field theory},
Lectures given at the Les Houches Summer School in Theoretical Physics,
Les Houches (France), June~28-August~5,~1988, 
e-print {\tt arXiv:hep-th/9108028} (August 1991).

\end{thebibliography}
\end{document}